\documentclass[prl,amssymb,twocolumn]{revtex4}
\usepackage{graphicx}
\begin{document}

\vskip 2mm

{\bf Comment on "Influence of Pair Breaking and Phase Fluctuations
on Disordered High $T_c$ Cuprate Superconductors"}

\vskip 2mm

In a recent Letter \cite{Rullier}, Rullier-Albenque {\it et al.} studied
the $T_c$ degradation under electron irradiation of Y-123
single crystals. They have measured the in-plane resistivity $\rho_{ab}(T)$
in a broad range of defect contents $x_d$,
the value of $x_d$ being proportional to $\Delta \rho_{ab}$, the increase in
$\rho_{ab}$ upon irradiation. It was found
that $T_c$ unexpectedly decreased quasilinearly with $x_d$ in the whole range
from $T_{c0}$ down to $T_c=0$.
The authors of Ref. \cite{Rullier} arrived
at a conclusion that experimental data are at variance with
Abrikosov-Gor'kov (AG) pair breaking theory \cite{AG} and
point to a significant role of phase fluctuations \cite{EK} of the
order parameter $\Delta({\bf p})$ in high-$T_c$ cuprates.
In this Comment, we show that the data reported in Ref. \cite{Rullier} are in
fact not inconsistent with the pair-breaking theories if (i) the deviation
from pure $d$-wave symmetry of $\Delta({\bf p})$ and (ii) the existence of
magnetic scatterers in irradiated samples are properly taken into account.

The authors of Ref. \cite{Rullier} made use of the AG formula \cite{AG}
for $d$-wave superconductors,
$\ln(T_{c0}/T_c)=\Psi(1/2+1/2\pi T_c\tau)-\Psi(1/2)$,
where $\tau$ is the electron
scattering time, $\tau^{-1}\propto x_d \propto \Delta \rho_{ab}$.
This formula gives a downward curvature of $T_c(\Delta \rho_{ab})$ curve,
contrary to experimental observations \cite{Rullier}.
Note, however, that the symmetry of
$\Delta({\bf p})$ in high-$T_c$ cuprates may be different
from pure $d$-wave \cite{Brandow,Zhao}. Besides, irradiation may result in
appearance of magnetic scatterers along with nonmagnetic ones since radiation
defects disturb antiferromagnetic correlations
between copper spins. The AG-like
formula that accounts for both those effects \cite{Openov} reads
$\ln(T_{c0}/T_c)=(1-\chi)\left[\Psi(1/2+1/2\pi T_c\tau_m)-\Psi(1/2)\right]+%
\chi\left[\Psi(1/2+1/4\pi T_c\tau_n+1/4\pi T_c\tau_m)-\Psi(1/2)\right]$,
where $\tau_n$ and $\tau_m$ are scattering times due to nonmagnetic and
magnetic defects, respectively, the coefficient $\chi=%
1-\langle\Delta({\bf p})\rangle^2_{FS}/\langle\Delta^2({\bf p})\rangle_{FS}$
is a measure of $\Delta({\bf p})$ anisotropy
on the Fermi surface ($\chi=1$ for $d$-wave, $0<\chi<1$ for mixed
$(d+s)$-wave or anisotropic $s$-wave, $\chi=0$ for isotropic $s$-wave).

An account for combined effect of both nonmagnetic and magnetic
scatterers on $T_c$ and/or an assumption about a non-pure $d$-wave
$\Delta({\bf p})$ allows for a quantitative explanation of the
experimental data within the AG-like pair breaking theory, without
resorting to phase fluctuations effects. Fig. 1 shows the measured
$T_c/T_{c0}$ versus $\Delta \rho_{ab}$ taken from Ref.
\cite{Rullier} along with the curves computed for $\chi=0.9$ and
various values of the coefficient
$\alpha=\tau_m^{-1}/(\tau_n^{-1}+\tau_m^{-1})$ that specifies the
relative contribution to the total scattering rate from magnetic
scatterers \cite{Openov}. Here we make use of the relation
\cite{Openov}
$\tau_n^{-1}+\tau_m^{-1}=(\omega_{pl}^2/4\pi)\Delta\rho_{ab}$,
where $\omega_{pl}$ is a characteristic energy which should not
necessarily coincide with the plasma frequency determined by, e.
g., optical spectroscopy. The quasilinear dependence of $T_c$ on
$\Delta\rho_{ab}$ in YBa$_2$Cu$_3$O$_7$ is quantitatively
reproduced at $\omega_{pl}=0.75$ eV and $\alpha=0\div 0.01$. This
value of $\omega_{pl}$ is a factor of 1.4 different from directly
measured values of the plasma frequency in Y-123. Although our
choice of $\omega_{pl}$ is, to some extent, arbitrary, the change
in $\omega_{pl}$ will result just in the change of the best fitting
values of $\chi$ and $\alpha$, e. g., $\chi\approx 0.8$ and 0.6,
$\alpha=0.04\pm 0.02$ and $0.04\pm 0.01$ at $\omega_{pl}=0.8$ and 1
eV, respectively. The data for YBa$_2$Cu$_3$O$_{6.6}$ can also be
well fitted within this approach.

\begin{figure}[h]
\includegraphics[width=\hsize]{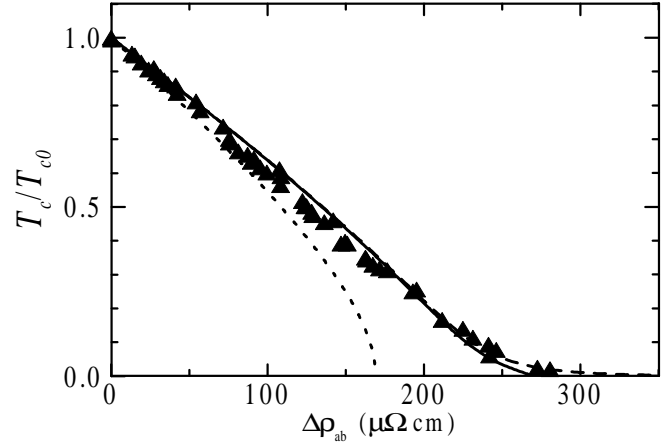}
\caption{$T_c/T_{c0}$ versus $\Delta\rho_{ab}$ in electron
irradiated YBa$_2$Cu$_3$O$_7$ crystals. Experiment \cite{Rullier}
(triangles). Theory \cite{Openov} for $\omega_{pl}=0.75$ eV,
$\chi=0.9$ and $\alpha=0$ (dashed line), 0.01 (solid line), and 1
(dotted line).} \label{Fig1}
\end{figure}

Finally, the arguments presented in Ref. \cite{Rullier} concerning the upward
curvature of $T_c(\Delta\rho_{ab})$ curve required to explain the maximum of
the transition width $\delta T_c$
as a function of $\Delta\rho_{ab}$ seem to be incompatible with experimental data
since the curvature of the measured $T_c(\Delta\rho_{ab})$ dependence is
close to zero in the whole range of $\Delta\rho_{ab}$.

\vskip 2mm

I am grateful to A. V. Kuznetsov for assistance.
The work was supported by the Russian Ministry of Industry,
Science, and Technology, Grant No 40.012.1.1.1357.

\vskip 2mm

L. A. Openov

Moscow Engineering Physics Institute, Moscow 115409, Russia.

\vskip 2mm


\vskip 2mm



\end{document}